\documentclass[onecolumn]{emulateapj}

\newcommand{\water}{H$_{2}$O }

\shorttitle{Optical spectroscopy of VUV irradiated pyrene:H$_2$O ice}
\shortauthors{Bouwman et al.}

\begin{document}

\title{{Real-time optical spectroscopy of VUV irradiated 
pyrene:H$_2$O interstellar ice}}

\author{J. Bouwman\altaffilmark{1}, D. M. Paardekooper\altaffilmark{1}, 
H. M. Cuppen\altaffilmark{1} and H. Linnartz\altaffilmark{1}}

\affil{Raymond \& Beverly Sackler Laboratory for Astrophysics, Leiden Observatory,\\ 
University of Leiden, NL-2300 RA Leiden, The Netherlands}

\and

\author{L. J. Allamandola\altaffilmark{2}}
\affil{NASA-Ames Research Center, Space Science Division,\\ Mail Stop 245-6,
Moffett Field, CA 94035}

\altaffiltext{1}{Raymond \& Beverly Sackler Laboratory for Astrophysics, Leiden Observatory, 
University of Leiden, NL-2300 RA Leiden, The Netherlands; bouwman@strw.leidenuniv.nl.}
\altaffiltext{2}{NASA-Ames Research Center, Space Science Division, Mail Stop 245-6,
Moffett Field, CA 94035.}

\begin{abstract}

This paper describes a near-UV/VIS study of a pyrene:H$_2$O 
interstellar ice analogue at 10 K using optical absorption
spectroscopy. A new experimental approach makes it possible to
irradiate the sample with vacuum ultraviolet (VUV) light 
(7-10.5 eV) while simultaneously recording spectra in the 
240-1000 nm range with subsecond time resolution. Both 
spectroscopic and dynamic information on VUV processed ices 
are obtained in this way. This provides a powerful tool to 
follow, \itshape in-situ \upshape and in real-time, the 
photophysical and photochemical processes induced by VUV 
irradiation of a polycyclic aromatic hydrocarbon containing inter- and circumstellar ice analogue. 
Results on the VUV photolysis of a prototype sample\textemdash 
strongly diluted pyrene in H$_2$O ice\textemdash are presented. 
In addition to the pyrene cation (Py$^+$), other products\textemdash hydroxypyrene (PyOH), possibly hydroxypyrene cation (PyOH$^+$), 
and pyrene/pyrenolate anion (Py$^-$/PyO$^-$)\textemdash 
are observed. It is found that the charge remains localized in the ice, also after the VUV irradiation is stopped. The astrochemical implications and observational constraints are discussed.

\end{abstract}


\section{Introduction}

At present, more than 150 different inter- and circumstellar 
molecules have been observed in space. The chemical diversity is
striking, and both simple and very complex  as well as stable and
transient species have been detected. Among these unambiguously
identified species polycyclic aromatic hydrocarbon molecules (PAHs) 
are lacking even though PAHs are generally thought to be ubiquitous 
in space \citep[e.g.,][]{vandishoeck04}. Strong infrared emission features
at 3.3, 6.2, 7.7, 8.6, and 11.2 {$\mu$}m are common in regions of,
for example, massive star formation and have been explained by PAH
emission upon electronic excitation by vacuum ultraviolet (VUV) radiation.
Consequently, PAHs are expected to play a key role in the heating of
neutral gas through the photoelectric effect. PAHs are also considered 
as important charge carriers inside dense molecular clouds, and
relevant for molecule formation through ion\textendash molecule interactions \citep{gillett73,puget89,allamandola89,kim01,smith07,draine07}.
Nevertheless, the only aromatic species unambiguously identified 
in space is benzene, following infrared observations 
\citep{cernicharo01}.

In recent years electronic transitions of PAH cations have been
studied in the gas phase with the goal to link laboratory data to
unidentified optical absorption features observed through diffuse
interstellar clouds. Following matrix isolation spectroscopic work 
\citep{salama91}, gas phase optical spectra have been recorded for
several PAH cations \citep{romanini99,brechignac99} by combining
sensitive spectroscopic techniques and special plasma expansions 
\citep{motylewski00,linnartz09}. Such optical spectra have unique
features and therefore provide a powerful tool for identifying PAHs
in space. So far, however, no overlap has been found between 
laboratory spectra of gaseous PAHs$^+$ and astronomical features.

In dense molecular clouds, most PAHs should quickly condense onto
the \water \textendash rich icy grain mantles, quenching the IR emission process.
Here, they will participate in ice grain chemistry. More than 25 years 
of dedicated studies, mainly in the infrared, have proven that a 
direct comparison between laboratory and astronomical ice spectra 
paints an accurate picture of the composition and the presence of 
inter- and circumstellar ices, even though solid-state features are 
rather broad. The spectral features (band position, band width 
(FWHM) and the intensity ratio of fundamental vibrations) depend 
strongly on mixing ratio and ice matrix conditions and this provides 
a sensitive analytical tool to identify ice compositions in space \citep[e.g.,][]{boogert08,oberg08}.

In the past, several experiments have been reported in which the 
formation of new molecules was proven upon VUV irradiation of 
astronomical ice mixtures, typically under high vacuum conditions 
\citep[e.g.,][]{mendoza95,bernstein99,gudipati03,ruiterkamp05,peeters05,elsila06}. 
Many of these studies were not \itshape in-situ\upshape, {\itshape i.e.}, reactants were
determined after warm up of the ice, and although VUV-induced
photochemistry at low temperatures is expected to take place, it is
not possible to fully exclude that at least some of the observed 
reactants may have been formed during the warm-up stage. 
More recently, \itshape in-situ \upshape studies have become
possible using ultra high vacuum setups in which ices are grown with
monolayer precision and reactions are monitored using reflection 
absorption infrared spectroscopy and temperature programmed desorption. 
Recent results show that H-atom bombardment of CO and O$_2$ ice results 
in the efficient formation of H$_2$CO/CH$_3$OH and H$_2$O$_2$/H$_2$O, respectively \citep{watanabe02,ioppolo08,miyauchi08}. Overall, such studies are still restricted to the
formation of rather small species, with ethanol as the most complex
molecule investigated in this way \citep{bisschop07}. Moreover, the reactants and products should not have overlapping bands in order to track them separately.

In the present work, a new approach is presented that extends our 
previous FTIR work on interstellar ice analogues to the UV/VIS. 
With the new setup it is possible to record, \itshape in-situ\upshape, 
the VUV photochemistry of PAHs and PAH derivatives in water ice at 10~K in 
real time. In the next section, this new approach is discussed in
detail. The first results for the PAH pyrene (C$_{16}$H$_{10}$, or Py) 
and its photoproducts, pyrene cation (C$_{16}$H$_{10}^+$, or 
Py$^+$), hydroxypyrene (PyOH), hydroxypyrene cation (PyOH$^{+}$),
and pyrene/pyrenolate anion (Py$^{-}$/PyO$^{-}$) in \water ice 
are presented in Section~\ref{spectroscopicassignment}. Finally, the astrophysical relevance of this work is discussed in Section~\ref{astrophysicalimplications}. The 
latter is twofold. First, time-dependent results
of VUV irradiated ice provide general insight into possible reaction
pathways upon photoprocessing of PAH containing water ice. Second, 
the results provide a spectroscopic alternative to search for PAH and
PAH-related optical features in the inter- and circumstellar medium
(ISM/CSM) through electronic solid-state absorptions.

\section{Experimental}
\label{experimental}

\begin{figure}[t]
\epsscale{1}
\plotone{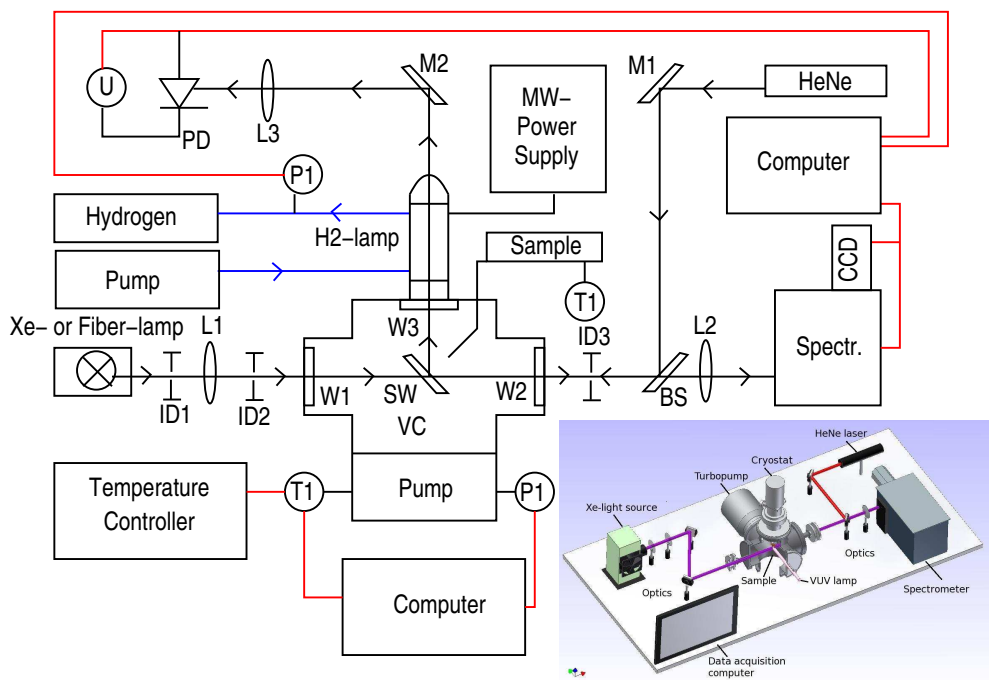}
\caption{A schematic drawing of the experimental setup. BS:~beam splitter; IDX:~iris diaphragm X; LX:~lens X; MX:~mirror X; PD:~photo diode to monitor interference fringes; PI:~pressure indicator; SW:~MgF$_2$ cold sample window; TI:~temperature indicator; U:~voltage meter; VC:~Vacuum Chamber and WX:~MgF$_2$ window X. The light paths are indicated by arrows, the data wiring is indicated in red lines, and the hydrogen flow is indicated by blue lines (see online color version). The inset shows a three-dimensional drawing of the experimental setup.
\label{fig1}}
\end{figure}

A schematic of the experimental setup is shown in Figure~\ref{fig1}. 
The experiment consists of three units: a vacuum chamber in which the
ice is grown, a special VUV irradiation source that is used for the
photo-processing of the ice and a source emitting broadband light that 
is focused into the ice and subsequently detected using a
monochromator equipped with a sensitive CCD camera.

The vacuum chamber consists of an ISO-160 6-cross piece. A 300~l~s$^{-1}$
turbomolecular pump, backed by a 10~m$^{3}$~hr$^{-1}$ double stage rotary
pump, is used to evacuate the chamber and to guarantee an operating
pressure of $\sim$10$^{-7}$~mbar. A catalytic trap is mounted
on the pre-vacuum pump to prevent pump oil from entering the vacuum
chamber.

The top flange of the cross piece holds a differentially pumped rotary flange on which is mounted a closed cycle helium refrigerator
equipped with a cold finger. A MgF$_2$ sample window with a diameter of
14 mm, clamped into an oxygen free copper holder between indium
gaskets, is mounted on the cold finger and centered on the optical axis
of the setup. This allows for rotation of the sample window through
360$^{\circ}$ under vacuum. The sample window can be cooled down to 
10 K and a thermocouple (Chromel-Au/Fe (0.07\%)) and a temperature
controller guarantee accurate temperature settings with 0.1~K
precision.

\begin{figure}
\epsscale{1}
\plotone{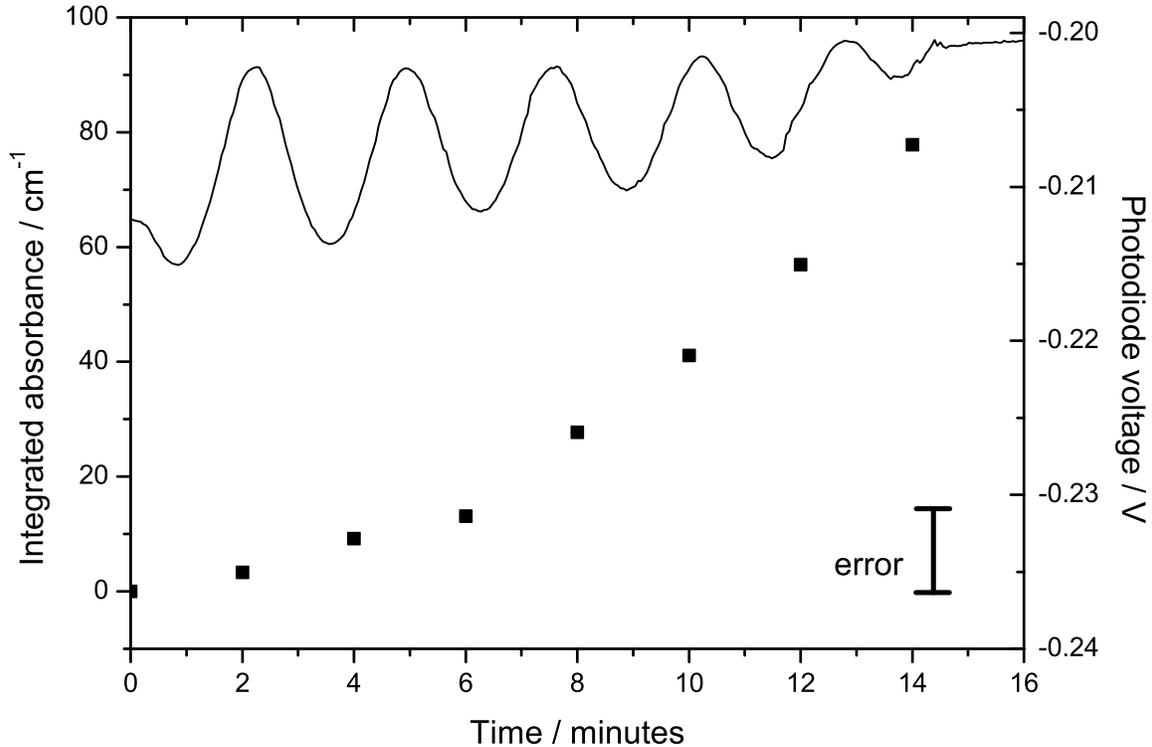}
\caption{A plot showing both the growth of the integrated pyrene absorption band (squares and left axis) and the interference fringes measured by the photodiode (right axis). At $t=14$ minutes, the deposition is stopped and the interference pattern diminishes. The error bar shown in the right lower corner applies to the pyrene-integrated absorbance.
\label{fig2}}
\end{figure}

The Py:H$_2$O sample is prepared by vapor depositing pyrene from a solid sample (Aldrich 99\%) heated to 40$^{\circ}$C, together with milli-Q water vapor from a liquid sample. The entire inlet system is maintained at $\geq$40$^{\circ}$C during deposition and comprises gas bulbs containing the sample material and tubing for directed deposition, approximately 15~mm from the sample window. The flow rate of the sample material is set by a high precision dosing valve. Condensation inside the tube is prohibited by additional resistive heating and the temperature settings are monitored by K-type thermocouples. The resulting ice film thickness is accurately measured by recording the number of interference fringe maxima ($m$) of a HeNe laser (${\lambda}$ =632.8~nm) which strikes the sample window at an angle of ${\theta}$=45$^{\circ}$. To monitor film growth and thickness, the intensity of the reflected laser light is measured with a sensitive photodiode. The ice thickness is subsequently determined by:
\begin{equation}
d=\frac{m\lambda_{\rm{HeNe}}}{2n_{{\rm{ice}}}\cos\theta},
\end{equation}
with the refractive index of the predominantly H$_2$O ice being $n_{\rm ice}\approx 1.3$ \citep{hudgins93}. This is illustrated in Figure~\ref{fig2} where both interference fringes produced during ice deposition and the simultaneous growth of the integrated neutral Py absorbance band are shown. The final ice thickness amounts to 1.7~$\mu$m and is reproducible to within 5\% or better. Simultaneously, the number of pyrene molecules in the ice sample ($N$) is monitored by measuring the integrated absorbance of its strongest transition (S$_2$ $\leftarrow$ S$_0$) (see Figure~\ref{fig2}). The number of pyrene molecules per cm$^{2}$ can be calculated via \citep{kjaergaard00,hudgins93}:
\begin{equation}
N=\frac{\int_{\nu_1}^{\nu_2} \tau\, d\nu}{8.88\times 10^{-13}f},
\end{equation}\label{eqn1}
where $f=0.33$ is the known oscillator strength of the S$_2$ $\leftarrow$ S$_0$ transition of pyrene \citep{bito00,wang03}. The resulting column density of pyrene molecules amounts to about 4$\times$10$^{14}$~cm$^{-2}$. For a typical sample with a thickness of 1.7~$\mu$m, the column density of H$_2$O molecules amounts to 4$\times$10$^{18}$~cm$^{-2}$, using the value for the density of amorphous ice ($\rho$=0.94~g/cm$^3$; \citet{sceats82}). Thus, the sample is calculated to consist of a 1:10,000 pyrene:H$_2$O mixture. This mixing ratio can be roughly varied by changing the H$_2$O flowrate or the pyrene sample temperature. The HeNe beam used here for monitoring the ice growth process also traces other elements along the optical path and is used to align all components.

The vacuum UV radiation from a special microwave (MW) powered hydrogen 
discharge lamp is used to simulate the interstellar radiation field
\citep{munoz02}. The lamp consists of a flow tube clamped in a 
McCarroll cavity \citep{mccarroll70} and emits mainly Ly-$\alpha$ 
radiation around 121.6~nm and, with less intensity, a band centered 
around 160~nm. The cavity is excited by a regular MW power supply 
(100~W, 2450~MHz). The H$_2$ pressure in the lamp is maintained at 
0.4~mbar during operation (Praxair 5.0 H$_2$). This results in a 
VUV photon flux of $\sim$10$^{15}$~photons~cm$^{-2}$~s$^{-1}$. 
The lamp is centered onto the front flange and the VUV radiation 
enters the setup towards the ice sample through a MgF$_2$ window that also serves as a vacuum seal. A shutter is used to block the VUV until the moment that 
the ice processing should start. Besides eliminating the need to switch the H$_2$ lamp on and off during the course of an experiment, this allows the lamp to stabilize before irradiation starts. This is important when tracking photochemical behavior during extended periods of photolysis. 

A 300~W ozone-free Xe-arc lamp serves as a broad band white light
source to measure the spectral ice features in direct absorption. 
The lamp has a spectral energy distribution that covers the full detector range (200~nm~$<\lambda<$~2400~nm). Alternatively, the light from a halogen fiber lamp can be used when no UV coverage is desired. An optical system
consisting of lenses and diaphragms is used to guide the light beam
through a MgF$_2$ window along the optical axis\textemdash coinciding with the
pre-aligned HeNe beam\textemdash and crossing the ice sample at a 45$^{\circ}$
angle. Light that is not absorbed exits the vacuum chamber through
a second MgF$_2$ window after which it is focused onto the entrance 
slit of an ANDOR Shamrock spectrometer. The spectrometer 
is equipped with two interchangeable turrets which holds four gratings
in total (2400, 1200, 600 and 150~lines~mm$^{-1}$), allowing for a trade-off
between wavelength coverage and spectral resolution, depending on the
experimental needs. Since typical ice absorption bands exhibit a 
FWHM of 4-20~nm, most of the experiments are performed using the 
600~lines~mm$^{-1}$ grating, resulting in an accessible wavelength range of 
$\sim$140~nm.

The light is dispersed onto a very sensitive 1024$\times$256 pixel 
CCD camera with 16-bit digitization. The resulting signal is read
out in the vertical binning mode by a data acquisition computer. 
Spectra are taken in absorbance mode ($\tau$=-ln($I/I_0$)) with 
respect to a reference spectrum ($I_0$) taken directly after 
depositing the sample. Recording a single spectrum typically takes
about 5~ms and spectra are generally co-added to improve
the signal-to-noise ratio (S/N). In a typical experiment more than 1000 individual spectra 
are recorded and are reduced using LabView routines. Data reduction
consists of local linear baseline corrections, multiple Gaussian
fitting of absorption profiles and absorption band integration.

\begin{figure}
\epsscale{1}
\plotone{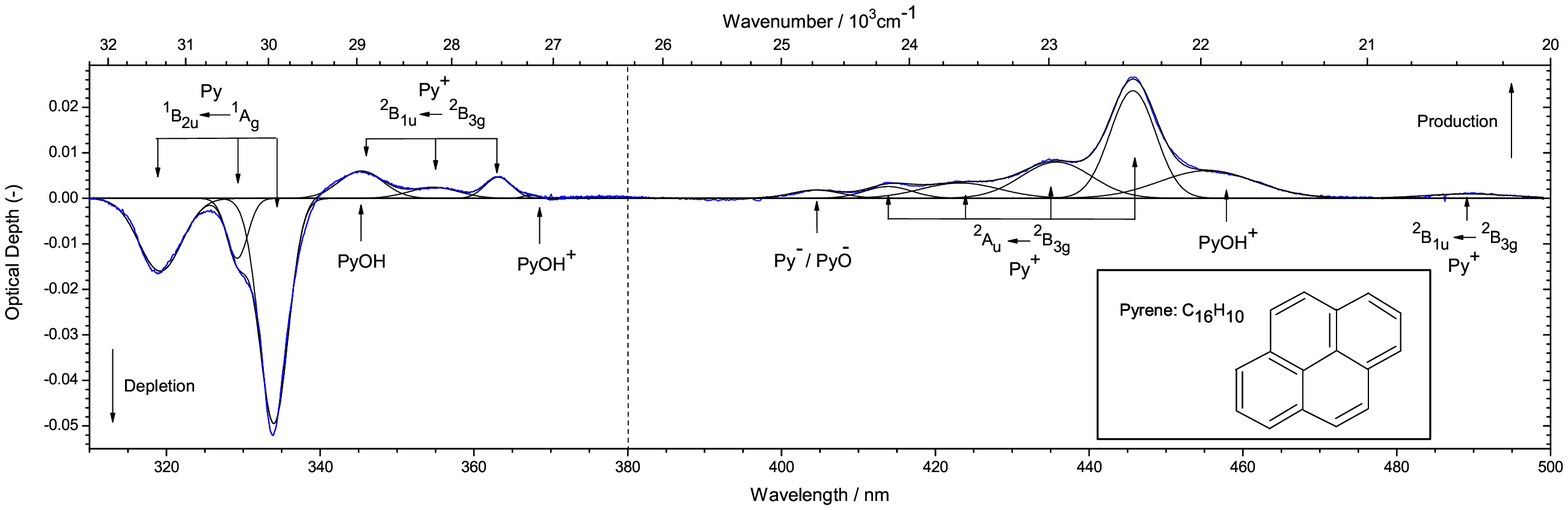}
\caption{A baseline corrected spectrum obtained after 1200 s of VUV irradiation of a pyrene:\water ice at 10 K. The absorption features in the plot are assigned and fitted with Gaussian profiles. A negative OD indicates species destruction, a positive OD species formation. The left spectrum (up to 280~nm) is scaled down by 30\% to facilitate comparison. 
\label{fig3}}
\end{figure}

\section{Spectroscopic assignment}
\label{spectroscopicassignment}

Figure~\ref{fig3} shows the 310 to 500~nm spectrum of a Py:H$_2$O ice at 10~K after 1200 s of \itshape {in-situ} \upshape photolysis in absorbance mode. The spectrum is baseline
subtracted and given in optical depth (OD). Since the spectrum recorded before VUV irradiation is taken as a reference ($I_{0}$), bands with positive OD values arise from species produced by photolysis while the carriers of negative OD bands decrease in density. It is noteworthy that S/N ratios are good even though the processes are studied in a very dilute mixture (Py:H$_2$O $\sim$1:10,000). Previous work was on more concentrated samples (PAH:H$_2$O $\sim$1:500, \citet{gudipati03}; PAH:H$_2$O $\sim$1:800 to 1:3200, \citet{bernstein99}).

A Gaussian fit to all of the features visible in the spectrum is indicated as well.
Clearly, a number of new species are produced at the expense of neutral pyrene. The peak positions, FWHM and assignments of all the bands in Figure~\ref{fig3} are summarized in Table~\ref{table1}, along with comparisons of earlier results found in other molecular environments. The assignments given in Figure~\ref{fig3} were made as follows.
Based on previous studies of pyrene in rare gas matrices 
\citep{vala94,halasinski05}, the strong, 
negative band peaking at 334~nm is readily assigned to the 
$^1B_{2u}$$\leftarrow$$^1A_g$ electronic transition of neutral pyrene ($S_2$$\leftarrow$$S_0$). Similarly, the positive bands near 363, 446, and 490~nm are assigned as the strongest members of the pyrene cation $^2B_{1u}$$\leftarrow$$^{2}B_{3g}$, $^2A_u$$\leftarrow$$^2B_{3g}$ and $^2B_{1u}$$\leftarrow$$^2B_{3g}$ vibronic transitions, respectively \citep{vala94,hirata99}. Table \ref{table1} shows that the bandwidth (FWHM) of the Py$^+$ bands is broader in the solid H$_2$O than in rare gas matrices, in accordance with the stronger interactions within the H$_2$O matrix network. 
Similarly, larger shifts in peak position may be expected. 

In addition to the Py$^+$ bands, other new bands appear near 345, 367, 405, and 453~nm. We ascribe these to hydroxypyrene (PyOH), hydroxypyrene cation (PyOH$^+$), and pyrene/pyrenolate anion (Py$^-$/PyO$^-$) based on the work of \citet{milosavljevic02} who reported the spectra of 1-hydroxypyrene and its daughter products in various media. The suggestion of anion production in these ices is noteworthy in view of the astronomical detection of negative ions both in the solid state \citep{broekhuizen05} and in the gas phase \citep[e.g.,][]{agundez08}.

The appearance of clear bands due to PyOH, PyOH$^+$, and 
Py$^-$/PyO$^-$ after VUV radiation at 10 K is somewhat surprising.  Previous optical studies of the VUV photolysis of three different PAHs in \water ice indicated that conversion of the parent neutral PAH to the cation was the major, and apparently only, photolytic step and that the cation remained stabilized in the ice to remarkably 
high temperatures ($\sim$100~K) for long periods \citep[e.g.,][]{gudipati06}. \citet{gudipati04} further showed that, in the case of naphthalene (Nap), subsequent reactions between Nap$^+$ and the 
\water matrix did indeed produce NapOH, but only during warm-up ($\sim$100~K).  Using mass spectroscopy, \citet{bernstein99} showed that PAH oxides and hydroxides were part of the residues left \itshape after \upshape VUV photolyzed PAH/H$_2$O ices were warmed under vacuum. Furthermore, upon prolonged exposure, the 
VUV transmittance of the MgF$_2$ hydrogen lamp window drops and with this 
the PyOH and PyOH$^+$ production also decreases relative to the production 
of Py$^+$.  This suggests that photolytic processes within the pyrene 
containing water ice change, perhaps because water dissociation becomes 
less effective with reduced hard UV flux while direct Py ionization 
still readily occurs with near UV photons \citep{gudipati04a}.  The influence 
of temperature and UV spectral energy distribution is currently under investigation.

\section{Chemical evolution of the ice}

\begin{figure}
\epsscale{1}
\plotone{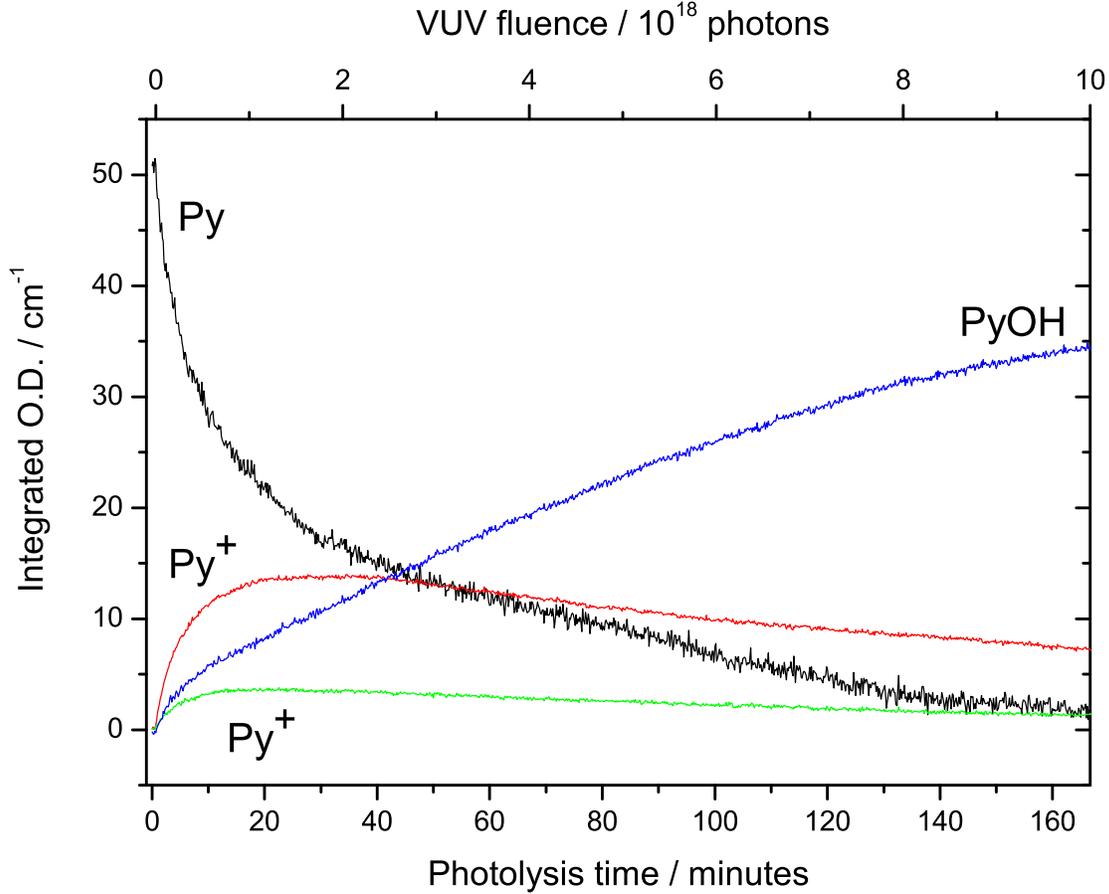}
\caption{The behavior of Py, Py$^+$ (two transitions), and PyOH as a function of VUV photolysis time/VUV fluence.
\label{fig4}}
\end{figure}

\begin{figure}
\epsscale{1}
\plotone{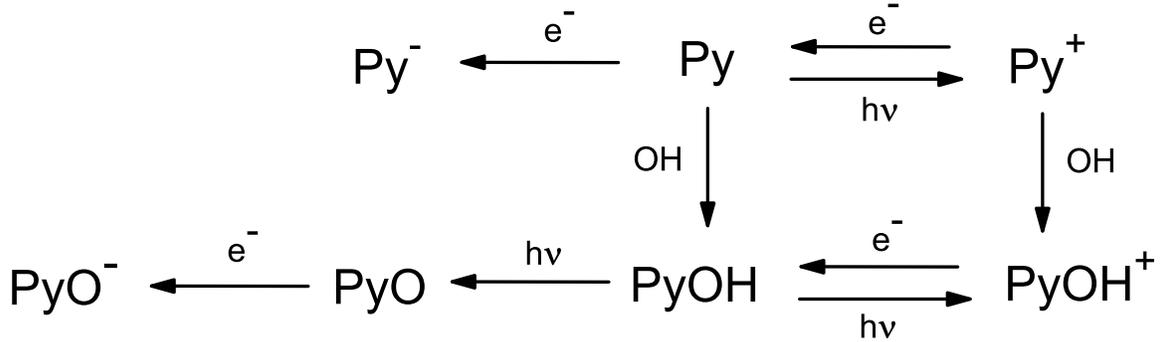}
\caption{Possible reactions upon photolysis of Py:H$_2$O ice as derived from Figure~\ref{fig3}.
\label{fig5}}
\end{figure}

To further investigate the spectroscopy and the photochemistry of VUV irradiated \water-rich ices that contain PAHs, time-dependent optical studies were performed. Figure~\ref{fig4} shows the integrated OD behavior of the Py, Py$^+$ (for two bands), and PyOH absorptions as function of photolysis time. During the first 130 s of VUV irradiation, the Py decay is clearly correlated with Py$^+$ growth. This allows us to determine relative band strengths of the two species by investigation of the short timescale correlation. The subsecond time response of the present setup is a prerequisite for this to work. We derive a band strength of $2.9\times 10^{-13}$ cm~molecule$^{-1}$ for the $^2A_{2u}\leftarrow ^2B_{3g}$ Py$^+$ transition in \water ice using Equation~(\ref{eqn1}). As other chemical processes become important, the correlation disappears. The loss of Py slows significantly while the Py$^+$ starts a slow decline after the maximum is reached. The PyOH signal continues to grow slowly but steadily throughout the photolysis process and is most likely formed by Py/Py$^+$ reacting with photoproducts of H$_2$O. Its formation is consistent with the recent outcome of a quantitative VUV photodesorption study of H$_2$O ice under ultra high vacuum conditions, where a H+OH photodissociation channel was reported \citep{oberg08a}. A reaction network connecting all of these species is presented in Figure~\ref{fig5}.

\begin{figure}
\epsscale{1}
\plotone{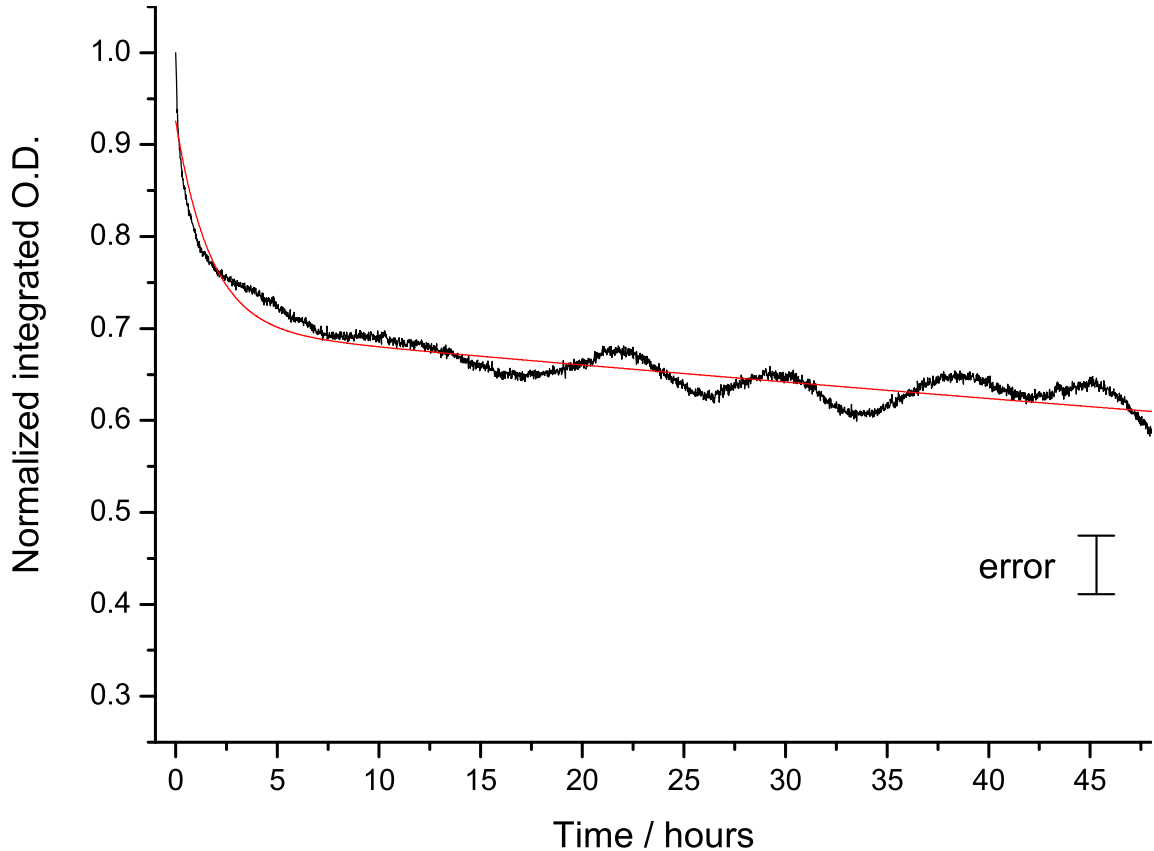}
\caption{Normalized pyrene cation integrated O.D. as a function of time plotted together with a double exponential fit using Equation~(\ref{exponenteqn}) (red curve). The wiggles superposed on the signal are caused by baseline effects and fall within the error of 5\% as indicated by the error bar.
\label{fig6}}
\end{figure}

In addition to Py$^+$ formation and reactivity during irradiation, we have also studied its stability within the ice when photolysis is stopped. Figure~\ref{fig6} plots the normalized integrated O.D. of the 445.6~nm Py$^+$ band as a function of time after VUV radiation is stopped. The figure spans 50~hr and shows that although the total Py$^+$ signal drops, 60\% remains trapped in the ice after 2 days. The small wiggle at the 0.05 level is due to baseline variations. There are clearly two decay channels, one \textquoteleft fast' and one \textquoteleft slow'. The following expression is used to fit the experimental data:
\begin{equation}
y=A_1\exp({-t/\tau_1})+A_2\exp({-t/\tau_2})
\label{exponenteqn}
\end{equation}
with $A_1=0.70$, $\tau_1=351.3$~hr, $A_2=0.22$ and $\tau_2=1.7$~hr. This produces the
red curve in Figure~\ref{fig6}. The processes responsible for these two decay rates are not yet clear. The Py$^+$ decay may be governed by recombination with trapped electrons. In considering these results, it is important to keep in mind that the ice processes described here are recorded for one temperature (10~K) and will most likely depend on temperature.

\begin{deluxetable}{l c c c c c c c c c c c}
\tablecaption{Vibronic bands of pyrene and its photoproducts in a \water ice compared with rare gas matrix literature values.\label{table1}}
\tabletypesize{\scriptsize}
\tablehead{\colhead{Species}&\colhead{}&\colhead{State}&\colhead{}&\colhead{$\lambda_{\rm{H_2O}}$} (nm)&\colhead{FWHM (nm)} &\colhead{}&\colhead{$\lambda_{\rm{lit.}}$ (nm)} &\colhead{FWHM$_{\rm{lit.}}$ (nm)}&\colhead{}&\colhead{$\lambda_{\rm{H_2O}}-\lambda_{\rm{rg}}$} &\colhead{$\Delta\rm{FWHM}$}}
\startdata
Py		&&$^1B_{2u}$& &334.0	&4.4	& &323.3\tablenotemark{a}	&n.a.	&	&10.7	&-\\
&&						& &329.2	&3.2	& &319.1\tablenotemark{a}	&n.a.	&	&10.1	&-\\
&&						& &319.2	&6.5	& &309.4\tablenotemark{a}	&n.a.	&	&9.8	&-\\
Py$^+$&&$^2B_{1u}$& &363.2	&3.6	& &362.6\tablenotemark{b}	&2.2	&	&0.6	&1.4\\
&	&					& &354.0	&6.5	& &355.1\tablenotemark{b}	&2.1	& 	&-1.1	&4.4\\
&	&					& &344.9	&6.2	& &-								&-		&	&-		&-\\
Py$^+$&&$^2A_u$	& &445.6	&6.6	& &443.8\tablenotemark{b}	&4.5	&	&1.8	&2.1\\
&	&					& &435.5	&10.2	& &433.2\tablenotemark{b}	&4.3	&	&2.3	&5.9\\
&	&					& &423.0	&12.2	& &422.9\tablenotemark{b}	&4.1	&	&0.1	&8.1\\
&	&					& &413.8	&5.3	& &412.1\tablenotemark{b}	&3.9	&	&1.7	&1.4\\
Py$^+$&&$^2B_{1u}$& &490.1	&10.0	& &486.9\tablenotemark{b}	&5.5	& 	&3.2	&4.5\\
\\
PyOH	&&				& &344.9	&5.8	& &340\tablenotemark{c}		&-		&	&-		&-\\
Py$^-$/PyO$^-$&&	& &405.2	&7.3	& &410\tablenotemark{c}		&-		&	&-		&-\\
PyOH$^+$		 & &	& &366.8	&3.0	& &-								&-		&	&-		&-\\
PyOH$^+$		& &	& &452.9	&18.2	& &465\tablenotemark{c}		&-		&	&-		&-
\enddata
\tablenotetext{a}{Values measured in a Ne matrix taken from \cite{halasinski05}}
\tablenotetext{b}{Values measured in an Ar matrix taken from \cite{vala94}}
\tablenotetext{c}{Values measured in \water and 2-chlorobutane taken from \cite{milosavljevic02}}
\end{deluxetable}

\section{Astrophysical implications}
\label{astrophysicalimplications}

Water is by far the dominant component of interstellar ices. 
Since PAHs are considered to be widespread throughout the
ISM, they are likely to be frozen out wherever H$_2$O-rich 
ices are present. The photochemical dynamics observed here
and the new spectroscopic information make two astrophysical points.

Astrochemically, this work shows that the effective photolytic production of PAH ions in PAH containing ice upon VUV irradiation should not be neglected {\it a priori} when modelling interstellar ice chemistry. The behavior of the various species that is shown in Figure~\ref{fig4} suggests that a new set of solid-state reactions appears when irradiating PAH containing water ice. The present study is on a rather isolated ice system\textemdash typical for this type of laboratory study\textemdash comprising Py and H$_2$O. In a more realistic interstellar sample, containing other constituents, such as CO, CO$_2$ or 
NH$_3$, and other PAHs, chemical pathways will become more complicated, but since water is the most dominant component in these extraterrestrial ices, we expect that the trends observed here will generally apply. Another important point to note is that reactions involving ions are not included in any of the current astrochemical grain chemistry networks. The present study shows that positive ions can reside in the ice mantle for a substantial time. This is particularly interesting since an astronomical dust grain is a truly isolated system, whereas the laboratory analogue is grown on the tip of a cold finger.

Observationally, the spectroscopic results provide an alternative route to search for PAH features in space. Astrophysical searches to identify PAHs and PAH cations in the ISM/CSM have focused on vibrational and
electronic transitions in the gas phase as well as solid-state 
PAH features in the infrared. As stated previously, these 
searches have been largely unsuccessful. The infrared work
suffers from spectral congestion and spectral overlap of 
vibrational modes. This prohibits an unambiguous identification of an emission feature to a specific carrier. In the UV/VIS this
is partly overcome as electronic excitations are unique for different PAHs. However, electronic spectra of gas phase PAHs showed no overlap with absorption features recorded through diffuse interstellar clouds, presumably of too low column densities \citep{romanini99,brechignac99}. A different situation applies in the solid state; PAHs are refractory material and accumulate in time onto cold grains that offer a reservoir. Therefore, the specific embedding of PAHs in water ice as presented here provides an alternative starting point for an astronomical search. However, one has to realize that this idea has both pros and cons.

Table \ref{table1} shows that under the present experimental conditions the bands are rather broad. The FWHM of the Py$^+$ absorption feature at 445.6 nm is 66 {\AA}. From an observational point of view this has the advantage that profiles can be studied at medium resolution, but also goes with the challenge to correct very accurately for possible background signals and in the end spectral overlaps may still exist, e.g., with silicate or carbonaceous
features. Nevertheless, overlapping broad bands can contribute to the very broad structure (VBS) superposed on the interstellar extinction curve \citep{hayes73,vanbreda81,krelowski86} and simply may have been overlooked in the past.

As stated before, the electronic excitation energy is unique. This is good for selectivity, but bad for sensitivity, since spectral features of different PAHs do not add up (as in the infrared) and consequently an individual optical band strength is, in principle, directly determined by the actual abundance of one specific PAH in space. It is difficult, however, to predict the percentage of the total interstellar PAH population that might exist in the form of one specific PAH, e.g., pyrene. Nevertheless, electronic PAH transitions are typically 2-3 orders of magnitude stronger than their IR bands and from this point of view, we expect that PAH bands may be observable in the visible even though IR bands are barely discernible on the strong \water bands \citep{brooke99}. This is also reflected by the very good S/N ratios as visible in Figure~\ref{fig2} for a very diluted mixture. 
We expect PAHs to be sufficiently abundant in ices in regions of molecular clouds with A$_v\geq$8 to permit detection in the optical. Infrared ice bands have been detected along such lines of sight and the visible extinction is low enough to permit UV from the interstellar radiation field to process these ices. 

We have estimated the expected Py and Py$^+$ absorption band strength towards an example source, MWC297. This is an early-type B1.5V star with a well-characterized stellar spectrum, a $B$ magnitude of 14.34 and a 3.0~$\mu$m ice band with $\tau=0.04$. Using the high sensitivity of an 8~m class telescope (VLT) it is feasible to obtain, within a couple of hours, S/N $\geq$1000 spectra of such a highly extincted source in the wavelength range under investigation. To estimate the expected OD of a pyrene or pyrene cation ice absorption we use the standard relation:
\begin{equation}
N_{\rm H}/E(B-V)=5.8\times 10^{21}~{\rm atoms~cm}^{-2}~{\rm mag}^{-1}
\end{equation}
from \citet{bohlin78}. With $E(B-V)$ = 2.67 towards MWC297 \citep{Drew97} this results in $N_{\rm H}$ = 1.6$\times$10$^{22}$~cm$^{-2}$. Taking the total PAH abundance in clouds with respect to $n_{\rm H}$ to be $\sim$3$\times$10$^{-7}$, this results in a total PAH column density of 9.6$\times$10$^{15}$~cm$^{-2}$. Assuming that 1\% of the total PAHs in space is in the form of pyrene frozen out on grains and of this fraction up to 10\% is in its singly ionized state (Py$^+$), the total column density of Py$^+$ or Py toward MWC297 is estimated to range between 9.6$\times$10$^{12}$ and 8.64$\times$10$^{13}$~species/cm$^{2}$. The OD is defined as
\begin{equation}
\tau=\frac{NA}{\Delta\nu},
\end{equation}
with $N$ the column density of the absorbers, $A$ the integrated bandstrength and $\Delta\nu$ the FWHM. For a typical strong allowed vibronic transition, such as the $^1B_{2u}$$\leftarrow$$^{1}A_{g}$ Py and $^2A_u$$\leftarrow$$^{2}B_{3g}$ Py$^+$ transitions, we take $A_{\rm Py}=2.9\times 10^{-13}$~cm~molecule$^{-1}$ and $A_{\rm Py^+}=2.9\times 10^{-13}$~cm~radical$^{-1}$ in ice with a FWHM=400 cm$^{-1}$ and 300~cm$^{-1}$, respectively. This yields ODs of $ 0.01\leq \tau \leq 0.06$ for Py and Py$^+$, a range similar to that observed for ice bands.

\section{Conclusion}

This work presents the first results of a spectroscopic and photo-chemical study of pyrene in water ice upon VUV irradiation under astronomical conditions. Since the spectra are recorded in real time, it is possible to derive photochemical characteristics and to monitor a rich ion-mediated chemistry in the solid state. Such processes are yet to be considered in astrochemical models. Additionally, it is shown that the pyrene cations formed within the H$_2$O ice by VUV irradiation remain trapped in the ice for an extended period. Successive heating of the ice makes these ions available to diffusing species and hence should be considered in solid-state astrochemical processes.

The new laboratory approach presented here offers a general way to
provide astronomically relevant PAH solid-state spectra. Specifically,
the spectra discussed here provide an alternative way to search for pyrene
features in the ISM/CSM. The derived numbers show the potential of this method, but one has to realize, as pointed out before, that these numbers
incorporate our limited knowledge on the actual PAH quantities in space.
For different PAHs, with different abundances and different absorption
strengths other numbers, both less and more favorable, may be expected. Furthermore, it is possible\textemdash in view of the rather effective way in which charged
species form and stay in the ice\textemdash that the actual abundance of ions may be
higher. The results presented here are new and aim at a further characterization of the chemical role of PAHs and PAH derivatives in space.

\acknowledgments

This work is financially supported by \textquoteleft Stichting voor
Fundamenteel Onderzoek der Materie\textquoteright (FOM), \textquoteleft the Netherlands Research School for Astronomy\textquoteright (NOVA) and NASA's Astrobiology Program. L. J. Allamandola thanks the \textquoteleft Nederlandse Organisatie voor Wetenschappelijk Onderzoek\textquoteright (NWO) for a visitors grant and NASA's Laboratory Astrophysics and Astrobiology Programsm for support.


\begin{thebibliography}{48}
\expandafter\ifx\csname natexlab\endcsname\relax\def\natexlab#1{#1}\fi

\bibitem[{{Ag{\'u}ndez} {et~al.}(2008){Ag{\'u}ndez}, {Cernicharo},
  {Gu{\'e}lin}, {Gerin}, {McCarthy}, \& {Thaddeus}}]{agundez08}
{Ag{\'u}ndez}, M., {Cernicharo}, J., {Gu{\'e}lin}, M., {et~al.} 2008, \aap,
  478, L19

\bibitem[{{Allamandola} {et~al.}(1989){Allamandola}, {Tielens}, \&
  {Barker}}]{allamandola89}
{Allamandola}, L.~J., {Tielens}, A.~G.~G.~M., \& {Barker}, J.~R. 1989, \apjs,
  71, 733

\bibitem[{{Bernstein} {et~al.}(1999){Bernstein}, {Sandford}, {Allamandola},
  {Gillette}, {Clemett}, \& {Zare}}]{bernstein99}
{Bernstein}, M.~P., {Sandford}, S.~A., {Allamandola}, L.~J., {et~al.} 1999,
  Science, 283, 1135

\bibitem[{{Bisschop} {et~al.}(2007){Bisschop}, {Fuchs}, {van Dishoeck}, \&
  {Linnartz}}]{bisschop07}
{Bisschop}, S.~E., {Fuchs}, G.~W., {van Dishoeck}, E.~F., \& {Linnartz}, H.
  2007, \aap, 474, 1061

\bibitem[{Bito {et~al.}(2000)Bito, Shida, \& Toru}]{bito00}
Bito, Y., Shida, N., \& Toru, T. 2000, Chem. Phys. Lett., 328, 310

\bibitem[{{Bohlin} {et~al.}(1978){Bohlin}, {Savage}, \& {Drake}}]{bohlin78}
{Bohlin}, R.~C., {Savage}, B.~D., \& {Drake}, J.~F. 1978, \apj, 224, 132

\bibitem[{{Boogert} {et~al.}(2008){Boogert}, {Pontoppidan}, {Knez}, {Lahuis},
  {Kessler-Silacci}, {van Dishoeck}, {Blake}, {Augereau}, {Bisschop},
  {Bottinelli}, {Brooke}, {Brown}, {Crapsi}, {Evans}, {Fraser}, {Geers},
  {Huard}, {J{\o}rgensen}, {{\"O}berg}, {Allen}, {Harvey}, {Koerner}, {Mundy},
  {Padgett}, {Sargent}, \& {Stapelfeldt}}]{boogert08}
{Boogert}, A.~C.~A., {Pontoppidan}, K.~M., {Knez}, C., {et~al.} 2008, \apj,
  678, 985

\bibitem[{{Br{\'e}chignac} \& {Pino}(1999)}]{brechignac99}
{Br{\'e}chignac}, P. \& {Pino}, T. 1999, \aap, 343, L49

\bibitem[{{Brooke} {et~al.}(1999){Brooke}, {Sellgren}, \& {Geballe}}]{brooke99}
{Brooke}, T.~Y., {Sellgren}, K., \& {Geballe}, T.~R. 1999, \apj, 517, 883

\bibitem[{{Cernicharo} {et~al.}(2001){Cernicharo}, {Heras}, {Tielens}, {Pardo},
  {Herpin}, {Gu{\'e}lin}, \& {Waters}}]{cernicharo01}
{Cernicharo}, J., {Heras}, A.~M., {Tielens}, A.~G.~G.~M., {et~al.} 2001, \apjl,
  546, L123

\bibitem[{{Draine} {et~al.}(2007){Draine}, {Dale}, {Bendo}, {Gordon}, {Smith},
  {Armus}, {Engelbracht}, {Helou}, {Kennicutt}, {Li}, {Roussel}, {Walter},
  {Calzetti}, {Moustakas}, {Murphy}, {Rieke}, {Bot}, {Hollenbach}, {Sheth}, \&
  {Teplitz}}]{draine07}
{Draine}, B.~T., {Dale}, D.~A., {Bendo}, G., {et~al.} 2007, \apj, 663, 866

\bibitem[{{Drew} {et~al.}(1997){Drew}, {Busfield}, {Hoare}, {Murdoch}, {Nixon},
  \& {Oudmaijer}}]{Drew97}
{Drew}, J.~E., {Busfield}, G., {Hoare}, M.~G., {et~al.} 1997, \mnras, 286, 538

\bibitem[{{Elsila} {et~al.}(2006){Elsila}, {Hammond}, {Bernstein}, {Sandford},
  \& {Zare}}]{elsila06}
{Elsila}, J.~E., {Hammond}, M.~R., {Bernstein}, M.~P., {Sandford}, S.~A., \&
  {Zare}, R.~N. 2006, Meteoritics and Planetary Science, 41, 785

\bibitem[{{Gillett} {et~al.}(1973){Gillett}, {Forrest}, \&
  {Merrill}}]{gillett73}
{Gillett}, F.~C., {Forrest}, W.~J., \& {Merrill}, K.~M. 1973, \apj, 183, 87

\bibitem[{Gudipati(2004)}]{gudipati04}
Gudipati, M. 2004, J.~Phys.~Chem.~A., 108, 4412

\bibitem[{{Gudipati} \& {Allamandola}(2003)}]{gudipati03}
{Gudipati}, M.~S. \& {Allamandola}, L.~J. 2003, \apjl, 596, L195

\bibitem[{{Gudipati} \& {Allamandola}(2004)}]{gudipati04a}
{Gudipati}, M.~S. \& {Allamandola}, L.~J. 2004, \apjl, 615, L177

\bibitem[{{Gudipati} \& {Allamandola}(2006)}]{gudipati06}
{Gudipati}, M.~S. \& {Allamandola}, L.~J. 2006, \apj, 638, 286

\bibitem[{{Halasinski} {et~al.}(2005){Halasinski}, {Salama}, \&
  {Allamandola}}]{halasinski05}
{Halasinski}, T.~M., {Salama}, F., \& {Allamandola}, L.~J. 2005, \apj, 628, 555

\bibitem[{{Hayes} {et~al.}(1973){Hayes}, {Mavko}, {Radick}, {Rex}, \&
  {Greenberg}}]{hayes73}
{Hayes}, D.~S., {Mavko}, G.~E., {Radick}, R.~R., {Rex}, K.~H., \& {Greenberg},
  J.~M. 1973, in IAU Symposium, Vol.~52, Interstellar Dust and Related Topics,
  ed. J.~M. {Greenberg} \& H.~C. {van de Hulst}, 83--+

\bibitem[{Hirata {et~al.}(1999)Hirata, Lee, \& Head-Gordon}]{hirata99}
Hirata, S., Lee, T., \& Head-Gordon, M. 1999, {J. Chem. Phys.}, 111, 8904

\bibitem[{{Hudgins} {et~al.}(1993){Hudgins}, {Sandford}, {Allamandola}, \&
  {Tielens}}]{hudgins93}
{Hudgins}, D.~M., {Sandford}, S.~A., {Allamandola}, L.~J., \& {Tielens},
  A.~G.~G.~M. 1993, \apjs, 86, 713

\bibitem[{{Ioppolo} {et~al.}(2008){Ioppolo}, {Cuppen}, {Romanzin}, {van
  Dishoeck}, \& {Linnartz}}]{ioppolo08}
{Ioppolo}, S., {Cuppen}, H.~M., {Romanzin}, C., {van Dishoeck}, E.~F., \&
  {Linnartz}, H. 2008, \apj, 686, 1474

\bibitem[{Kim {et~al.}(2001)Kim, Wagner, \& Saykally}]{kim01}
Kim, H.-S., Wagner, D.~R., \& Saykally, R.~J. 2001, Phys. Rev. Lett., 86, 5691

\bibitem[{{Kjaergaard} {et~al.}(2000){Kjaergaard}, {Robinson}, \&
  {Brooking}}]{kjaergaard00}
{Kjaergaard}, H.~G., {Robinson}, T.~W., \& {Brooking}, K.~A. 2000, J. Phys.
  Chem. A, 104, 11297

\bibitem[{{Krelowski} {et~al.}(1986){Krelowski}, {Maszkowski}, \&
  {Strobel}}]{krelowski86}
{Krelowski}, J., {Maszkowski}, R., \& {Strobel}, A. 1986, \aap, 166, 271

\bibitem[{{Linnartz}(2009)}]{linnartz09}
{Linnartz}, H. 2009, Cavity ring-down spectroscopy; Techniques and
  applications, ed. G.~{Berden} \& R.~{Engeln} (Wiley), in press

\bibitem[{McCarroll(1970)}]{mccarroll70}
McCarroll, B. 1970, Rev. Scient. Instrum., 41, 279

\bibitem[{{Mendoza-Gomez} {et~al.}(1995){Mendoza-Gomez}, {de Groot}, \&
  {Greenberg}}]{mendoza95}
{Mendoza-Gomez}, C.~X., {de Groot}, M.~S., \& {Greenberg}, J.~M. 1995, \aap,
  295, 479

\bibitem[{Milosavljevic \& Thomas(2002)}]{milosavljevic02}
Milosavljevic, B. \& Thomas, J. 2002, Photochem. Photobiol. Sci., 1, 100

\bibitem[{{Miyauchi} {et~al.}(2008){Miyauchi}, {Hidaka}, {Chigai}, {Nagaoka},
  {Watanabe}, \& {Kouchi}}]{miyauchi08}
{Miyauchi}, N., {Hidaka}, H., {Chigai}, T., {et~al.} 2008, Chemical Physics
  Letters, 456, 27

\bibitem[{{Motylewski} {et~al.}(2000){Motylewski}, {Linnartz}, {Vaizert},
  {Maier}, {Galazutdinov}, {Musaev}, {Kre{\l}owski}, {Walker}, \&
  {Bohlender}}]{motylewski00}
{Motylewski}, T., {Linnartz}, H., {Vaizert}, O., {et~al.} 2000, \apj, 531, 312

\bibitem[{{Mu{\~n}oz Caro} {et~al.}(2002){Mu{\~n}oz Caro}, {Meierhenrich},
  {Schutte}, {Barbier}, {Arcones Segovia}, {Rosenbauer}, {Thiemann}, {Brack},
  \& {Greenberg}}]{munoz02}
{Mu{\~n}oz Caro}, G.~M., {Meierhenrich}, U.~J., {Schutte}, W.~A., {et~al.}
  2002, \nat, 416, 403

\bibitem[{{{\"O}berg} {et~al.}(2008){{\"O}berg}, {Boogert}, {Pontoppidan},
  {Blake}, {Evans}, {Lahuis}, \& {van Dishoeck}}]{oberg08}
{{\"O}berg}, K.~I., {Boogert}, A.~C.~A., {Pontoppidan}, K.~M., {et~al.} 2008,
  \apj, 678, 1032

\bibitem[{{{\"O}berg} {et~al.}(2009){{\"O}berg}, {Linnartz}, {Visser}, \& {van
  Dishoeck}}]{oberg08a}
{{\"O}berg}, K.~I., {Linnartz}, H., {Visser}, R., \& {van Dishoeck}, E.~F.
  2009, \apj, 693, 1209

\bibitem[{{Peeters} {et~al.}(2005){Peeters}, {Botta}, {Charnley}, {Kisiel},
  {Kuan}, \& {Ehrenfreund}}]{peeters05}
{Peeters}, Z., {Botta}, O., {Charnley}, S.~B., {et~al.} 2005, \aap, 433, 583

\bibitem[{{Puget} \& {Leger}(1989)}]{puget89}
{Puget}, J.~L. \& {Leger}, A. 1989, \araa, 27, 161

\bibitem[{{Romanini} {et~al.}(1999){Romanini}, {Biennier}, {Salama},
  {Kachanov}, {Allamandola}, \& {Stoeckel}}]{romanini99}
{Romanini}, D., {Biennier}, L., {Salama}, F., {et~al.} 1999, Chem. Phys. Lett.,
  303, 165

\bibitem[{{Ruiterkamp} {et~al.}(2005){Ruiterkamp}, {Peeters}, {Moore},
  {Hudson}, \& {Ehrenfreund}}]{ruiterkamp05}
{Ruiterkamp}, R., {Peeters}, Z., {Moore}, M.~H., {Hudson}, R.~L., \&
  {Ehrenfreund}, P. 2005, \aap, 440, 391

\bibitem[{{Salama} \& {Allamandola}(1991)}]{salama91}
{Salama}, F. \& {Allamandola}, L.~J. 1991, \jcp, 94, 6964

\bibitem[{{{Sceats}, M.~G. and {Rice}, S.,~A.}(1982)}]{sceats82}
{{Sceats}, M.~G. and {Rice}, S.,~A.} 1982, Water: A Comprehensive Treatise,
  Vol.~7 (Plenum, New York), 83--214

\bibitem[{{Smith} {et~al.}(2007){Smith}, {Draine}, {Dale}, {Moustakas},
  {Kennicutt}, {Helou}, {Armus}, {Roussel}, {Sheth}, {Bendo}, {Buckalew},
  {Calzetti}, {Engelbracht}, {Gordon}, {Hollenbach}, {Li}, {Malhotra},
  {Murphy}, \& {Walter}}]{smith07}
{Smith}, J.~D.~T., {Draine}, B.~T., {Dale}, D.~A., {et~al.} 2007, \apj, 656,
  770

\bibitem[{{Vala} {et~al.}(1994){Vala}, {Szczepanski}, {Pauzat}, {Parisel},
  {Talbi}, \& {Ellinger}}]{vala94}
{Vala}, M., {Szczepanski}, J., {Pauzat}, F., {et~al.} 1994, J. Phys. Chem., 98,
  9187

\bibitem[{{van Breda} \& {Whittet}(1981)}]{vanbreda81}
{van Breda}, I.~G. \& {Whittet}, D.~C.~B. 1981, \mnras, 195, 79

\bibitem[{{van Broekhuizen} {et~al.}(2005){van Broekhuizen}, {Pontoppidan},
  {Fraser}, \& {van Dishoeck}}]{broekhuizen05}
{van Broekhuizen}, F.~A., {Pontoppidan}, K.~M., {Fraser}, H.~J., \& {van
  Dishoeck}, E.~F. 2005, \aap, 441, 249

\bibitem[{{van Dishoeck}(2004)}]{vandishoeck04}
{van Dishoeck}, E.~F. 2004, \araa, 42, 119

\bibitem[{{Wang} {et~al.}(2003){Wang}, {Chang}, {Tso}, {Hsu}, \&
  {Cheng}}]{wang03}
{Wang}, B.~C., {Chang}, J.~C., {Tso}, H.~C., {Hsu}, H.~F., \& {Cheng}, C.~Y.
  2003, J. Mol. Struct. (Theochem), 629, 11

\bibitem[{{Watanabe} \& {Kouchi}(2002)}]{watanabe02}
{Watanabe}, N. \& {Kouchi}, A. 2002, \apjl, 571, L173

\end{thebibliography}



\end{document}